\documentclass[smallextended]{svjour3}       % onecolumn (second format)

\smartqed  % flush right qed marks, e.g. at end of proof

\usepackage{graphicx}
\usepackage{natbib}
\usepackage{graphicx}
\usepackage{amssymb}
\usepackage{amsfonts}
\usepackage{amsmath}
\usepackage{latexsym}
\usepackage{theorem}
\usepackage{paralist}
\usepackage{textcomp}
\usepackage{xspace}
\usepackage{subfigure}
\usepackage{graphicx}
\usepackage{xspace}
\usepackage{geometry}
\usepackage{microtype}
\newcommand{\U}{\mathbf{U}}
\newcommand{\x}{\mathbf{x}}
\newcommand{\ui}{\mathbf{u}}
\newcommand{\rx}{\mathbf{r}}
\newcommand{\rt}{\mathbf{r_T}}

\begin{document}
%\preprint{AIP/123-QED}

\title{Homogeneity and isotropy in a laboratory turbulent flow}% Force line breaks with \\
\titlerunning{ Homogeneity and isotropy in a laboratory turbulent flow}
\author{Gabriele Bellani \and Evan A. Variano}
\authorrunning{G.~Bellani \and E.~A.~Variano}

\institute{G. Bellani \at
              Department of Civil and Environmental Engineering, University of California, Berkeley,   CA 94720, USA\\
              \email{bellani@berkeley.edu}
            \and
            E. A. Variano \at
              Department of Civil and Environmental Engineering, University of California, Berkeley, CA 94720, USA}
\date{Received: date / Accepted: date}
\maketitle 

\begin{abstract}
We present a new design for a stirred tank that is forced by two parallel planar arrays of randomly actuated synthetic jets.  This arrangement creates turbulence at high Reynolds number with low mean flow.  Most importantly, it exhibits a region of 3D homogeneous isotropic turbulence that is significantly larger than the integral lengthscale.  These features are essential for enabling laboratory measurements of turbulent suspensions.  We use quantitative imaging to confirm isotropy at large, small, and intermediate scales by examining one-- and two--point statistics at the tank center.  We then repeat these same measurements to confirm that the values measured at the tank center are constant over a large homogeneous region.  In the direction normal to the symmetry plane, our measurements demonstrate that the homogeneous region extends for at least twice the integral length scale $L=9.5$ cm.  In the directions parallel to the symmetry plane, the region is at least four times the integral lengthscale, and the extent in this direction is limited only by the size of the tank.  Within the homogeneous isotropic region, we measure a turbulent kinetic energy of $6.07 \times 10^{-4} $m$^2$s$^{-2}$, a dissipation rate of $4.65 \times 10^{-5} $m$^2$s$^{-3}$, and a Taylor--scale Reynolds number of $R_\lambda=334$.  
%Having such a large homogeneous region, high Reynolds number, and low mean flow makes this stirred tank an optimal facility for studying the fundamental dynamics of turbulence and turbulent suspensions.  ALT: 
The tank's large homogeneous region, combined with its high Reynolds number and its very low mean flow, provides the best approximation of homogeneous isotropic turbulence realized in a laboratory flow to date.  These characteristics make the stirred tank an optimal facility for studying the fundamental dynamics of turbulence and turbulent suspensions.

\end{abstract}

%\keywords{Turbulence, Isotropic, Forcing, Particles}%Use showkeys class option if keyword
                              %display desired

\section{Introduction}

Homogeneous isotropic turbulence (HIT) is an idealized flow  of special interest because it contains all of the basic physical processes of turbulence without the complications commonly found in nature such as mean shear, density stratification, and fluid--solid boundaries \citep{Tsinober:2004p17930}. 
Thus HIT is an ideal flow with which to understand some  of the fundamental mechanisms of turbulence that are at least qualitatively independent of the origin of a specific turbulent flow such as: internal intermittency \citep{Douady:1991p23522}; the self--amplification mechanisms of velocity derivatives \citep{Galanti:2000p23995}; inertial range Eulerian and Lagrangian structure functions \citep{Benzi:2010p23252}; and (of particular interest to us) interphase coupling mechanisms in turbulent suspensions \citep{Poelma:2006p15458,Lucci:2010p15399,Balachandar:2010p13976,Toschi:2009p13919}.

Despite the simplicity of HIT, it is non--trivial to recreate this condition in a laboratory experiment or in a direct numerical simulation (DNS).  In DNS, turbulence either decays with time or must be sustained via an artificial forcing in space or time that  introduce biases in the turbulent statistics \citep{Abdelsamie:2012p24119,Lucci:2010p15399}.
Turbulent flows in laboratory experiments, on the other hand, are intrinsically inhomogeneous because it is impossible in practice to uniformly distribute turbulent production. 
At best, laboratory devices can only approximate HIT.  In doing  so, there has typically been a tradeoff between Reynolds number  and the size of the HIT region.  Herein, we present a new design for a stirred tank that achieves an unprecedented combination of size and Reynolds number, and conduct a thorough characterization.

\section{Background}

The most common way of generating turbulence for laboratory research is with a steady flow passing through a grid or mesh.  These flows exhibit 2D homogeneity and isotropy in planes parallel to the grid \citep[see for example][and references therein]{Kurian:2009p24110,Krogstad:2012p23266}. Despite the good planar homogeneity, grid--generated turbulence is always anisotropic due to the spatial decay of turbulent kinetic energy (TKE) downstream of the grid.  This can make it difficult to compare results between different experimental setups. In fact, the large scatter in decay exponents and empirical coefficients suggests that there may not be a universal state for grid turbulence \citep{George:1992p24570,George:2004p24571}.

Stationary turbulence with 3D homogeneity and isotropy is produced by a new class of laboratory devices that have flourished in the past decade.  These devices stir the flow from multiple locations rather than with a single grid.  Stirring is conducted by means of oscillating grids \citep{Srdic:1996p23573,Shy:1997p23720,Villermaux:1995p23911},  loudspeaker cones \citep{Hwang:2004p23470,Birouk:2003p23787}, rotating elements \citep{Liu:1999p23761,Guala:2008p14427,Voth:2002p14090}, or synthetic jets \citep{Variano:2004p23918,Krawczynski:2010p26179,Goepfert:2010p26074}.
An essential feature of these systems is that the stirring elements are arranged symmetrically around some central region. This achieves large--scale isotropy, which in turn fosters small--scale isotropy.  
Of these symmetric forcing (SF) systems, the most common employ spherically symmetric forcing (SSF).  Early implementations of SSF used eight synthetic jets or fans at the corners of a box \citep[see][respectively]{Hwang:2004p23470, Birouk:2003p23787}.  Extensions of this idea have added more forcing elements and distributed them symmetrically over polygons with more than eight vertices \citep{Chang2012,zimmerman2010}. %Chang:2012p26016,

A drawback of SSF systems is that the flow is optimized only in a limited volume around the point of symmetry.  
Cylindrical Symmetric Forcing (CSF) and Planar Symmetric Forcing (PSF) systems allow larger regions of optimal flow conditions because they have a line or plane of symmetry at the tank center.  As a result, the optimal region at the tank center has at least one direction in which its homogeneity is limited only by the size of the tank.  
Because of this, we conclude that PSF systems  unite the best aspects of grid turbulence and SF systems.  That is, 2D homogeneity and isotropy are present throughout the tank due to the planar forcing, and 3D homogeneity and isotropy are present in a subregion of the tank due to the interaction of two symmetric forcing planes.  
Herein we present a PSF system that uses two planar arrays of randomly actuated synthetic jets.  

In table \ref{table:tank} we summarize a subset of the stirred tanks reported in the literature \citep{Srdic:1996p23573,Shy:1997p23720,Villermaux:1995p23911,Hwang:2004p23470,Birouk:2003p23787,Liu:1999p23761,Guala:2008p14427,Voth:2002p14090,Goepfert:2010p26074,zimmerman2010}.  Since there are many versions of the SSF cube, we report only a few representative cases; a more thorough comparative summary is given in \cite{Chang2012}.  We observe in table \ref{table:tank} a trend in which the flows with large Reynolds number have a small region of HIT, and vice versa.   Thus different stirred tanks will be optimal for different research needs. For our research on particle dynamics, we would like a high Reynolds number ($R_{\lambda}>>100$) and a region of homogeneous turbulence that is significantly larger than the integral lengthscale.  None of the devices in table \ref{table:tank} provide both of these traits simultaneously, though those of \cite{zimmerman2010} and \cite{Srdic:1996p23573} are closest.  Herein, we present a new device that provides high--Reynolds--number turbulence that is homogeneous and isotropic over a large region.  

The device we present herein can support a wide variety of investigations into turbulent dynamics.  An illustrative example is the measurement of macroscopic particles suspended in turbulent flows.  When performing such measurements \citep{BellaniJFM2012}, we require a flow with features that are not found in any other existing device (\emph{e.g.} table \ref{table:tank}).  First, a high Reynolds number is needed to ensure an inertial subrange.  Second, the integral lengthscale must be large enough that the inertial subrange will cover the size range of our macroscopic particles (5-30 mm).  Having an inertial subrange at such large scales also helps us make unambiguous measurements of small--scale turbulent features \citep[see \emph{e.g.}][]{superpipe}.  The dynamics of particles in suspension demand that we engineer a turbulent flow that is homogeneous and isotropic over as large a region as possible.  This is because the equation of motion for suspended drops, bubbles, and particles includes a history term \citep[\emph{e.g.}][]{Mei:1992p25755}.  This means that kinematics measured at one location implicitly include the integrated effect of the flow experienced over a particle's recent trajectory.  The flow facility presented here allows us to be more confident that particle kinematics measured at the tank center will represent the effects of only one type of turbulence, because the particles must travel through a large region of homogeneous isotropic turbulence before reaching the tank center.  Furthermore, the tank presented herein provides a distinct advantage for studies of buoyant particles: because the tank symmetry is not spherical, the homogeneous isotropic region can be extended indefinitely in the vertical direction, so that the test particles will not rise or fall through the homogeneous region too rapidly.

 \begin{table} 
\centering
\begin{tabular}{cccccc} 
\hline
\hline
\emph{Symmtery}     &  $\frac{r_{HIT}}{L} $ & $R_{\lambda}$ & $r_{HIT} (cm)$ & $L (cm)$  & \emph{Reference} \\

cylindrical  & 0.4 & 290 &	2.0 &4.7	&	 \cite{Liu:1999p23761}\\
spherical    & 0.5 & 480 &	5.0 &9.9	&	 \cite{Chang2012} \\
spherical    & 0.6 & 220 &	1.8 &2.8	&	 \cite{Hwang:2004p23470}\\
spherical    & 0.7 & 240 & 	2.5 &3.6	&	 \cite{Goepfert:2010p26074} \\
spherical    & 1.0 & 195 &	4.8 &4.7	&	 \cite{zimmerman2010}\\
spherical    & 1.2 & 92  &	2.0 &1.7	&	\cite{Birouk:2003p23787}	\\
planar       & 4.0 & 150  &	8.8&2.2	&	\cite{Srdic:1996p23573}	\\
planar       & 5.0  & 30  &	1.5 &0.3	&	 \cite{Shy:1997p23720}	\\
&&&&\\
planar      & 1.0 & 340 &	9.5 &  9.5  	&	This study	\\
\hline
\hline
 \end{tabular}
 \caption{A summary of stirred tank performance, focusing on the Reynolds number and size of the homogeneous region.  $r_{HIT}$ is the radius of the spherical region over which the flow is homogeneous and isotropic, or for rectangular regions, the half-width of the shortest dimension.  $L$ is the longitudinal integral lengthscale in this region.  $L$ is computed from autocorrelation functions or from $L \approx  u'^3 / 2 \epsilon$; comparisons of the two methods show excellent agreement \citep{Variano:2008p12422,zimmerman2010}.  In the case of \cite{Birouk:2003p23787} we have converted their reported transverse lengthscale to a longitudinal integral lengthscale using the factor of 2 predicted for isotropic turbulence.  The majority of citations define the homogeneous region as that over which turbulent kinetic energy varies by less than 10\%, and we follow this convention in this table, though this leads to slight underestimates of $r_{HIT}$ for three studies \citep{Liu:1999p23761,Chang2012,Srdic:1996p23573}.}  
 \label{table:tank}
\end{table}

\section{Experimental setup} 
\subsection{Description of the facility}

The experimental facility is a tank of dimensions 80$\times$80$\times$360 cm$^3$. The origin of the coordinate system is at the center of the tank, $z$ is oriented along the axial (longest) dimension of the tank, and $y$ is vertical. The instantaneous velocity vector $\mathbf{U}(x,y,z,t)=(U, V, W)$ is defined so that $U, V$ and $W$ are aligned with the $x$, $y$ and $z$ axes, respectively. The tank is filled with tap water, which is initially filtered to 5 micron and purified by a flow--through ultraviolet filter when experiments are not being run.  

\begin{figure}
\centering
\includegraphics[width=0.9\textwidth]{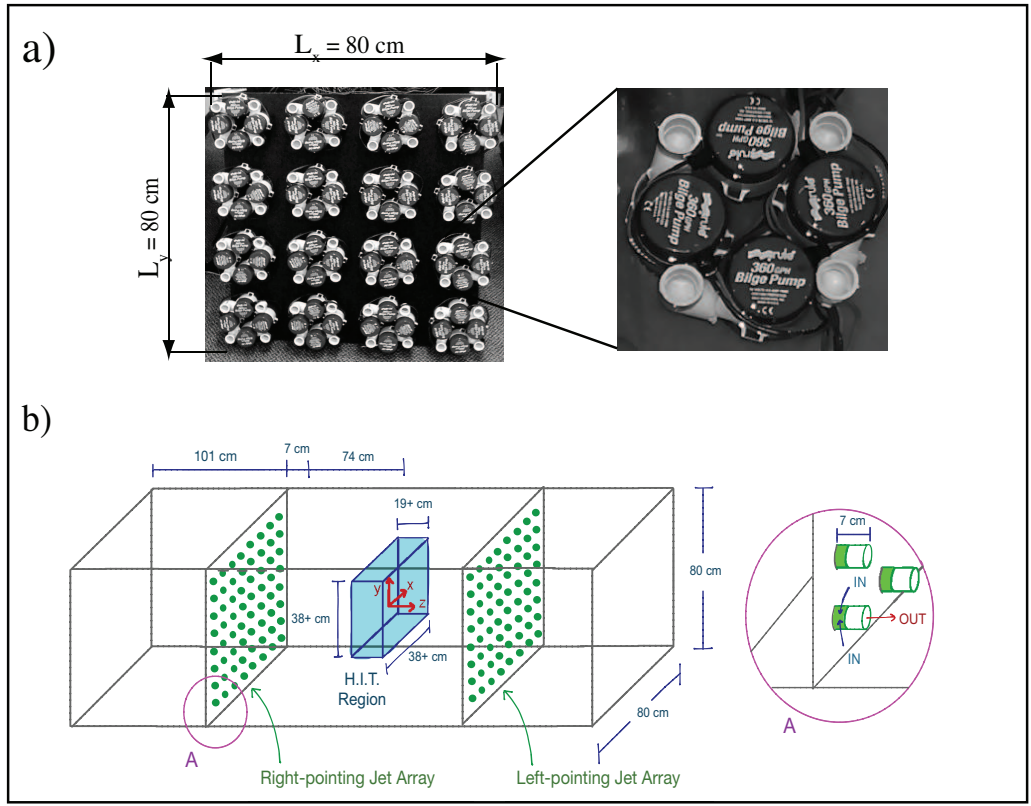}\\
\caption{Experimental facility. (a) 4$\times$4 array of 4-jet clusters (see closeup), giving an 8$\times$8 array of equally spaced jets. Two pumps arrays facing each other (as in sketch b) can produce homogeneous and isotropic turbulence statistics.}\label{fig:tank_setup}  
\end{figure}

Stirring is provided by two facing planes of randomly actuated synthetic jet arrays, each of them made of 64 individual pumps arranged in an 8$\times$8 array as shown in figure~\ref{fig:tank_setup}. Each pump is used to create a jet through a cylindrical nozzle with 2.19 cm inner diameter.  The nozzle and the pump intake are separated by 7 cm and located in the same volume of fluid; thus the pump creates a ÒsyntheticÓ jet, in the sense that it injects only momentum, and not mass, into the tank. 
To drive the pumps, we follow the stochastic algorithm developed by \cite{Variano:2008p12422}. The algorithm is a stochastic pattern used to drive the jets in a manner that maximizes the turbulent Reynolds number while also ensuring spatial homogeneity. In this algorithm, an average of eight jets (12.5\% of the total number of jets) are activated on each planar array, with each jet remaining actuated for an average duration, $\mu_{on}$, of 3 s, followed by an average time turned off, $\mu_{off}$, of 21s. The exact duration of a jet's on/off period is selected from normal probability distributions, where the variance value of each distribution ($\sigma_{on}$;$\sigma_{off}$)  is such that $\sigma_{on}/\mu_{on}=\sigma_{off} /\mu_{off}$=1/3. These values were found, experimentally, to maximize Reynolds number by maximizing shear production of turbulence. The stochasticity of the algorithm prevents any tank--scale residual flow from persisting; having negligible mean flow serves experiments in two ways.  First, it supports homogeneity and isotropy by reducing advective transport of TKE.  Second, it allows the spatial and temporal characteristics of turbulence to be measured independently from a fixed location.  Thus measurements do not need to rely on Taylor's frozen turbulence hypothesis, allowing a less ambiguous analysis of turbulent structures \citep{Dennis:2008p12424,DelAlamo:2009p12425,Moin:2009p12426}. 

Turbulence is generated near the jet arrays and decays with distance from them. By combining two arrays in a PSF configuration it is possible to obtain HIT in a large region in the tank center. The two jet arrays are symmetrically located with respect to the  vertical center--plane of the tank, at a distance of $\pm$81 cm from the center.  This distance is chosen to maximize the isotropy at the tank center by matching the decay curves of the 3 different components of the velocity variance so that they intersect at the tank center.

%The details of the flow will be discussed further in section \ref{sec:singphas}.

\subsection{Measurement technique}
\label{sec:meas}
\begin{figure}
\centering
\includegraphics[width=0.99\textwidth]{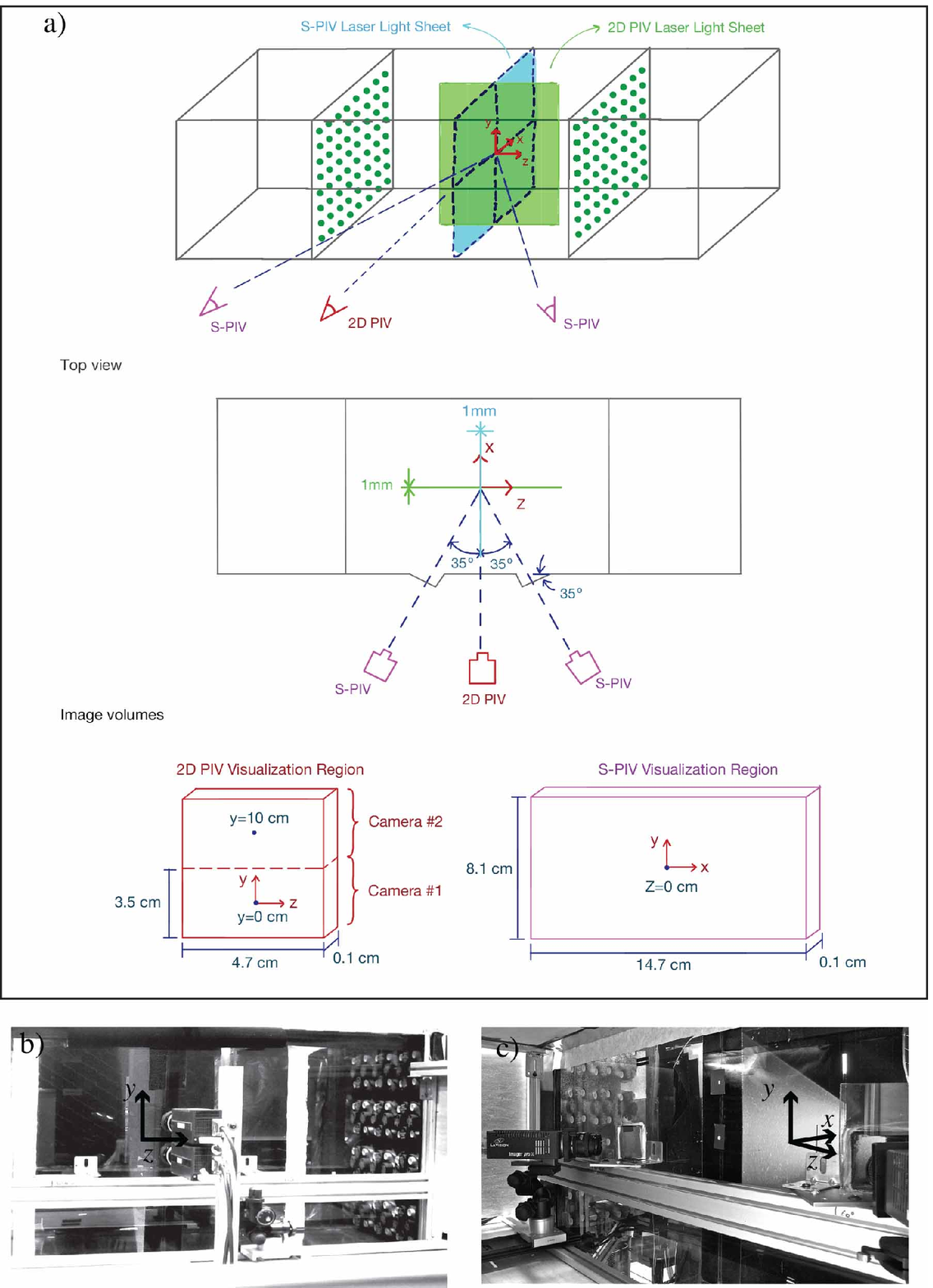} 
\caption{a) Schematic of the imaging setup for 2D-PIV  and S-PIV measurements. Pictures of the 2D-PIV and S-PIV setups are shown in b) and c), respectively.}\label{fig:piv_setup}  
\end{figure}

Velocity measurements are performed using Particle Image Velocimetry (PIV) in two different configurations (see figure \ref{fig:piv_setup}).  2D-PIV is used to collect data in the $y-z$ plane and Stereoscopic PIV (S-PIV) is used to collect data in the $x-y$ plane.  The imaging setups are shown in figure \ref{fig:piv_setup}.  Both PIV configurations use a 1mm thick laser light sheet (frequency-doubled Nd-YAG), 10 $\mu \textrm{m}$ tracer particles (silver coated glass spheres), two 12-bit CCD cameras with an 1600 $\times$ 1200 array of 7.4 $\mu$m pixels (Imager PRO-X), image-pair acquisition rate of 0.5 Hz, and either 50 mm or 105 mm lenses (Nikkor).  In the 2-D PIV setup, two cameras (both fitted with a 105 mm Nikkor lens) view two adjacent regions, thus increasing the extent of spatial coverage. These regions are 0.1 cm $\times$ 3.5 cm $\times$ 4.7 volumes centered in the $y-z$ plane at $y=0$ and $y=10$ cm, respectively. In the S-PIV experiments, the two cameras, both fitted with a 105 mm Nikkor lens, view one measurement area from opposite sides of the laser light sheet, each at an angle of 35 degrees relative to the laser's forward-scatter direction. To avoid image distortion by the air-glass-water interface at the tank walls, two 35$^\circ$ prisms filled with water are attached to the walls. Each camera views the tank through one prism, and images overlap in a 14.7 cm $\times$ 8.1 $\times$ cm $\times$ 0.1 cm  volume centered in the $x-y$ plane at $z=0$. 

%Stereoscopic PIV applied to the images provides 2-dimensional arrays of 3-dimensional velocity vectors, with 61 $\times$ 109 measurement points equally spaced at 1.3635 mm grid size.  This grid spacing is approximately equal to the laser sheet thickness. 

To compute the velocity fields for both PIV configurations, we use the commercial software \emph{Davis 7.2} from \emph{LaVision GmbH}.  
The main PIV operating parameters are reported in table \ref{table:piv}. 
These parameters can greatly effect PIV accuracy in measuring turbulent quantities. The main sources of error in PIV measurements are well understood \citep{RaffelPIV}. Particular care is needed near boundaries and in high shear; only the latter is of concern here.  For this reason, we use an algorithm with continuous window deformation and reduction that has been shown to perform well in turbulence studies with objective benchmarks \citep{Stanislas:2008p9497}.  
%Another bias can arise from a finite spatial resolution. 
When spatial resolution is coarser than the Kolmogorov microscale, velocity fluctuations and TKE may be underestimated.  Of those spatial resolutions reported in table \ref{table:piv}, the coarsest is 2.72 mm.  This is about 7 times the size of the Kolmogorov scale, which is fine enough to resolve $>$95\% of the TKE \citep{Saarenrinne:2001gx}.   We explicitly confirm that we have resolved the TKE adequately by comparing the velocity fluctuation magnitudes as measured with two different resolutions (see figure~\ref{fig:urms3}). 

 \begin{table} 
\centering
\begin{tabular}{ccccc}                      
\hline
\hline
                         &        \emph{IA} [pixels $\times$ pixels]                &     \emph{IA}  [mm $\times$ mm]         & Weighting function &Vector spacing [mm]     \\
       S-PIV                 &       32 $\times$ 32                       &       2.72 $\times$ 2.72          & Gaussian & 1.36        \\
      2D-PIV                 &       32 $\times$ 32                        &               0.88 $\times$ 0.88     &   Square & 0.44       \\
\hline
\hline
 \end{tabular}
 \caption{Summary of PIV settings. The size of the final interrogation area (IA) and the weighting function determine the spatial resolution of the PIV measurements. The Gaussian weighting function used by \emph{Davis} is a symmetric 2D Gaussian window with $\sigma$ = [(IA/2)-1], whose radius at $e^{-2}$ of the peak amplitude is $r_v$= 2.64 mm in the case shown here.}\label{table:piv}
\end{table}

\section{Definitions}
Turbulent statistics are computed as follows.  Expectation values, denoted $\langle \cdot \rangle$, are estimated from appropriate space, time, or ensemble averages. 

\begin{tabular}{ll}
Coordinate system: &$\x=\{x, y, z\}=\{$lateral, vertical, axial$\}$\\
Velocity field: &$\U=\{U, V, W\}$.\\
Fluctuating field: & $\ui(\x,t)=\U(\x, t)-\langle{\U}(\x)\rangle$.\\
Turbulent Kinetic Energy (TKE): & $k^2=\frac{1}{2} \langle\mathbf{u}\cdot\mathbf{u}\rangle \approx \frac{1}{2}\langle 2v^2+w^2 \rangle$.\\
Longitudinal structure function:& $S_L^{(p)}(\rx)=\langle\{[\ui(\x+\rx)-\ui(\x)]\cdot \frac{\rx}{|\rx|}\}^p\rangle$.\\
Transverse structure function: & $S_T^{(p)}(\rx)=\langle\{[\ui(\x+\rt)-\ui(\x)]\}^p\rangle$, with $\rt \perp \ui$.\\
Longitudinal autocovariance:& $C_L(\rx)=\langle\{[\ui(\x+\rx) \cdot \ui(\x)]\cdot \frac{\rx}{|\rx|}\}^p\rangle$.\\
Transverse autocovariance: & $C_T(\rx)=\langle\{[\ui(\x+\rt) \cdot \ui(\x)]\}^p\rangle$, with $\rt \perp \ui$.
\end{tabular}

\section{Results}
\label{sec:Result}

\subsection{Homogeneity and isotropy of single point statistics}
\label{sec:one_point}
\begin{figure}[tbp]
\centering
\includegraphics[width=0.85\textwidth]{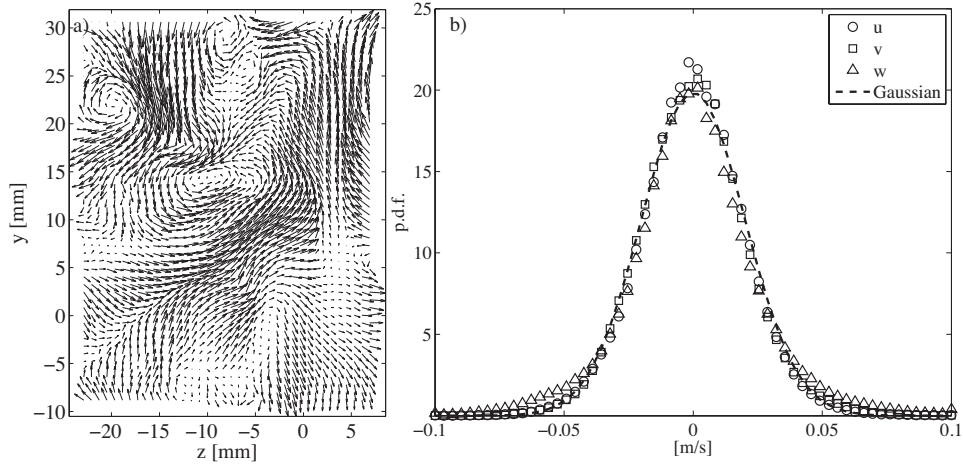}
\caption{a) Instantaneous velocity field showing $v$ and $w$ components in the $y-z$ plane measured by 2D-PIV. b) Probability distribution functions of the three velocity components from S-PIV measurements at the center of the tank  (1517 independent snapshots). The dashed line indicates a Gaussian distribution $z=\frac{1}{\sigma\sqrt{2\pi}} e^{-(\frac{x-\mu}{\sigma})^2}$, with $\sigma=v_{rms}$ and $\mu=0$.}\label{fig:piv}  
\end{figure} 

\begin{table} 
\centering
\begin{tabular}{ccccccc}                      
\hline
\hline
$\#$ \emph{Samples} &   & &  \emph{Mean} & \vbox{\hbox{\strut \emph{95\% CI}}\hbox{\strut \emph{(spatio-temporal)}}} & \vbox{\hbox{\strut \emph{95\% CI}}\hbox{\strut \emph{(over space)}}}& \vbox{\hbox{\strut \emph{95\% CI}}\hbox{\strut \emph{(over time)}}} \\
                      & $w_{rms}$ & [$\times10^{-2}$ ms$^{-1}$]  & 2.02  & [ 2.01    2.02 ]& [1.83 2.21]   &  [1.85 2.25] \\
       400         & $v_{rms}$ & [$\times10^{-2}$ ms$^{-1}$]  & 1.89   &[1.88 1.89] &[1.81 1.98]   &  [1.70 2.07] \\
                      & $k^2$ &  [$\times10^{-4}$ m$^2$s$^{-2}$]    & 5.74  & [5.71    5.78 ] &[5.17 6.31]   &  [5.16 6.29] \\
\\                      
                      & $w_{rms}$ & [$\times10^{-2}$ ms$^{-1}$]  & 2.01  &[ 2.00   2.01] &[1.85 2.17]   &  [1.90 2.19] \\
       750         & $v_{rms}$ & [$\times10^{-2}$ ms$^{-1}$]  & 1.93   & [1.93 1.94] &  [1.84 2.01]   &  [1.78 2.06] \\
                      & $k^2$ &  [$\times10^{-4}$ m$^2$s$^{-2}$]    & 5.79  & [5.75 5.82] &[5.26 6.32]   &  [5.40 6.24] \\
                      \\                      
                      & $w_{rms}$ & [$\times10^{-2}$ ms$^{-1}$]  & 2.08  &[ 2.08 2.09] &[1.92 2.25]   &  [1.97 2.23]   \\
       1000       & $v_{rms}$ & [$\times10^{-2}$ ms$^{-1}$]  & 1.98   & [1.98 1.99] &[1.89 2.06]   &  [1.86 2.10]     \\
                      & $k^2$ &  [$\times10^{-4}$ m$^2$s$^{-2}$]    & 6.07  & [ 6.03   6.10] &[5.50 6.64]   &  [5.77 6.54]   \\
\hline
\hline
 \end{tabular}
 \caption{Statistical convergence and spatial homogeneity, demonstrated using confidence intervals on velocity fluctuation magnitudes calculated over time or space. The temporal CI intervals are computed from $N_t$ samples in time (see first column), and 1 sample in space assuming a Gaussian distribution. The CI over space is the $95\%$ percentile variation of $100\times75$ samples in space. The spatio-temporal CI are determined from $100\times75\times N_t$ samples using the bootstrap method.}\label{table:res}
\end{table}

\begin{figure}[tbp]
\centering
\includegraphics[width=0.75\textwidth]{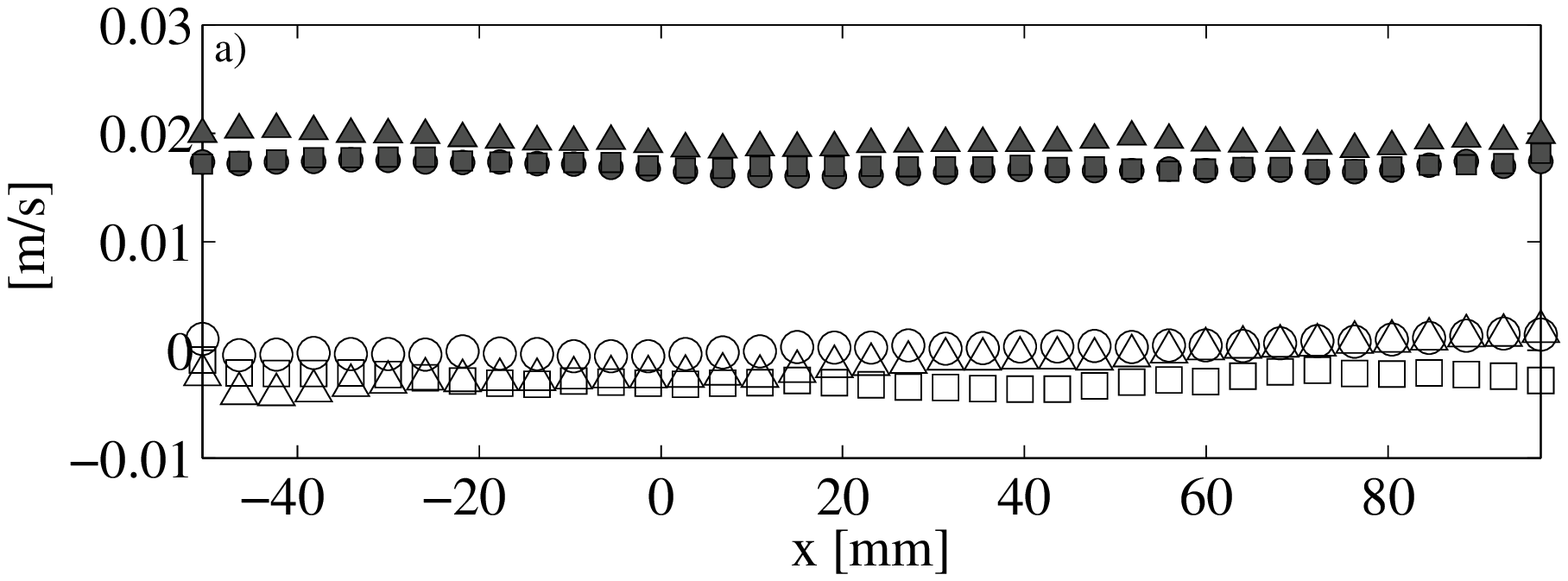}\\
\includegraphics[width=0.75\textwidth]{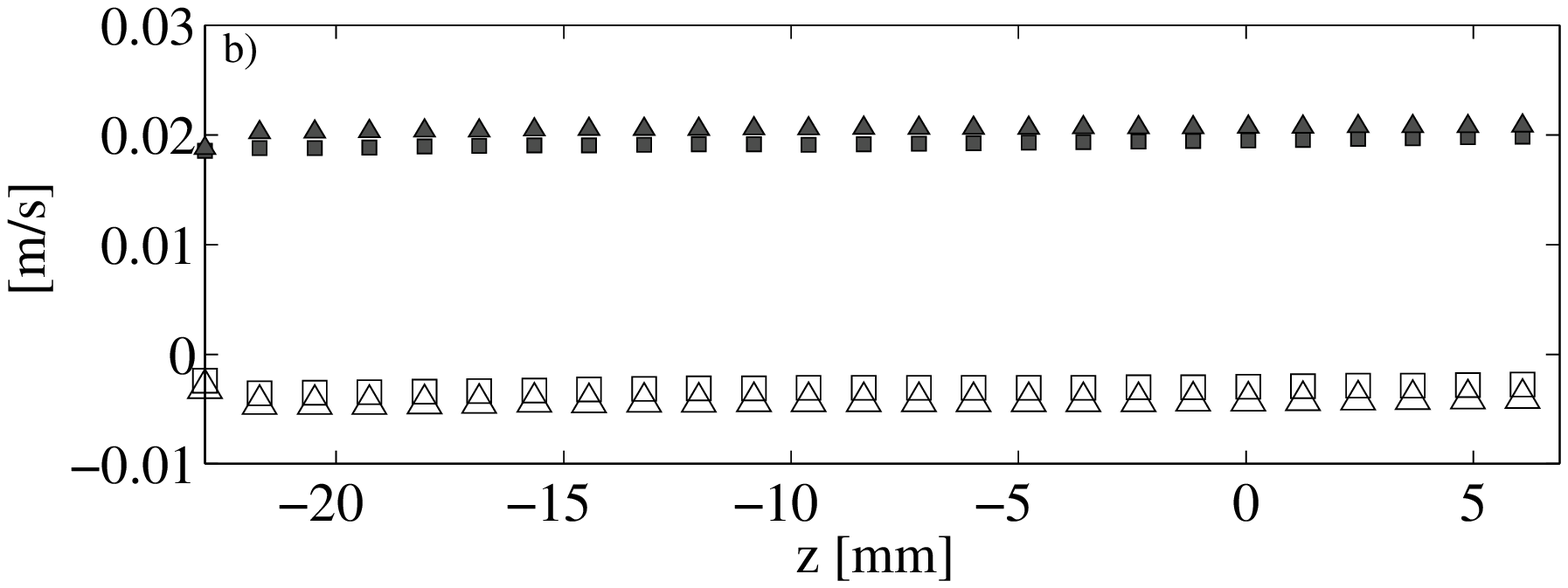}
\caption{a) one--point velocity statistics measured with S-PIV on the $x-y$ plane and averaged over $y$.  b) one--point velocity statistics measured with 2D-PIV on the y-z plane and averaged over y. In both plots, $\langle U \rangle$ ($\circ$), $\langle V \rangle$ ($\Box$), $\langle W \rangle$ ($\triangle$) and $u_{rms}$,$v_{rms}$,$w_{rms}$ are the corresponding filled markers. Marker sizes are scaled to show the 95\% confidence intervals. For clarity, only every third sample in space is shown here. Note different spatial scales on the two plots.}\label{fig:urms3}  
\end{figure}
 
At the tank center, statistics of turbulent fluctuations are isotropic, as seen in the velocity \emph{pdf}s in figure~\ref{fig:piv}.
The \emph{pdf}s of the three velocity components (calculated from 1517 S-PIV snapshots) are very similar, and they are very well approximated by a Gaussian distribution. 
 
Figure \ref{fig:urms3} shows the spatial distribution of mean velocities (with the mean determined over timeseries at each location).  It also shows the magnitude of velocity fluctuations (rms values computed from the temporal variance at each location). Data in Figure  \ref{fig:urms3}a comes from S-PIV measurements in the $x-y$ plane, while figure  \ref{fig:urms3}b shows 2D-PIV measurements in the $z-y$ plane. This data indicates that, at the center of our tank, the mean flow is negligible compared to the magnitude of velocity fluctuations.  A second conclusion from this data is that the mean and rms velocities are homogeneous over the center region of the tank; determining the full extent of the homogeneous region is the topic of section 6.  

A summary of the data from figure \ref{fig:urms3}b is given in table \ref{table:res}.  This includes a value of rms velocity calculated over both space and time.  The spatiotemporal dataset has 7300 spatial locations (covering the entire 2D PIV image area) and N temporal samples.  The N samples can be considered iid (independent and identically distributed) because they are recorded at 0.5 Hz, while the 7300 spatial locations are spaced too closely to be entirely independent from each other.  To evaluate the statistical convergence of the rms velocity value, we consider only the temporal data, because it is iid.  At one specific location, we calculate the 95\% CI on rms velocity using the bootstrap.  This interval is seen in Table \ref{table:res}, and it appears to be insensitive to increasing N above 400, thus indicating statistical convergence.   
Having obtained statistical convergence for rms velocity, we evaluate how the rms velocity varies in space.  We do so by calculating how rms velocity fluctuations (calculated from N temporal samples) vary across the 7300 spatial locations.  We quantify this spatial variation with a 95\% confidence interval computed with the bootstrap.  The resulting intervals (seen in table \ref{table:res}) overlap the temporal confidence intervals, and are similar in size.  This indicates that the variation over space is no larger than the uncertainty over time, and can be considered a quantitative statement of the spatial homogeneity observed in figure \ref{fig:urms3}.  

Figures \ref{fig:piv} and \ref{fig:urms3} and table \ref{table:res} allow us to evaluate isotropy at the tank center.  In figure \ref{fig:piv}, the \emph{pdf}s of $u$ and $v$ are nearly identical and the ratio between their standard deviations is $u_{rms}/v_{rms}\approx 1$. The maximum anisotropy and maximum deviation from the Gaussian curve is seen in $w$, which is to be expected because these motions are normal the symmetry plane.From the data in table \ref{table:res} the ratio $v_{rms}$/$w_{rms}$=0.95, with a 95\% confidence interval (CI) of [0.84 1.09] at any one location, computed via the bootstrap method \citep{efron_introduction_1994}.  If we compute the confidence interval of the isotropy ratio using the entire spatial extent of the 2D-PIV measurement at the tank center, it becomes [0.944 0.956].  Thus the velocity variance at the tank center is either within 10\% or 5\% of isotropy, depending on which confidence interval we choose to use. As variance primarily represents the large scale motions in turbulence, we conclude from these measurements that the flow at the tank center is isotropic at large scales.   This large--scale isotropy will promote isotropy at smaller scales, which is investigated below using two--point statistics. 

\subsection{Two-point statistics and turbulent scales}
\label{sec:two_point}
 
\begin{figure}[h!]
\centering
\includegraphics[width=0.75\textwidth]{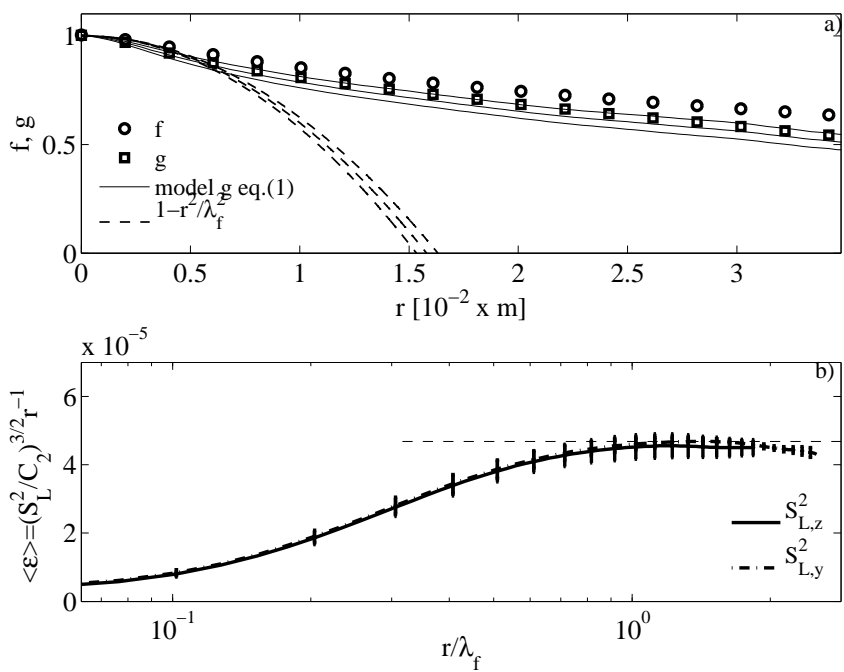}
\caption{a) Longitudinal ($\circ$) and transverse ($\square$) autocorrelation functions ($f(r)$ and $g(r)$, respectively). The size of the symbols is representative of the 95\% confidence interval (CI). The solid lines show the isotropic prediction of the transverse autocorrelation function $g(r)$ from eq.~\ref{eq:fg}, with its 95\% CI. The dashed line shows the parabolic fit used to compute the longitudinal Taylor lengthscale from the longitudinal autocorrelation and it 95\% CI.
b) Compensated second order structure functions used to determine dissipation rate (dashed line). The error bars represent 95\% CI.  The curves are obtained from 1517 independent 2D-PIV measurements.}\label{fig:SF2}  
\end{figure}

In HIT, the autocovariances $C_L$ and $C_T$ are isotropic, \emph{i.e.} independent of the direction of $r$. Normalizing both by their respective values at $r$=0 gives two  autocorrelation functions $f(r)$ and $g(r)$. 
The measurements of these longitudinal and lateral autocorrelation functions $f$ and $g$ are shown in figure \ref{fig:SF2}a.
Using the continuity equation as a constraint, $g$ can be expressed as a function of $f$ as shown by eq.~\ref{eq:fg}.
\begin{equation}\label{eq:fg}
g(r)=f(r)+\frac{1}{2}\frac{\partial f(r)}{\partial r}r.
\end{equation}
We can test isotropy by comparing the measured $g$ to the prediction obtained from equation~\ref{eq:fg}. The $g(r)$ predicted from equation (\ref{eq:fg}) is compared to a direct measurement of $g(r)$ in Figure \ref{fig:SF2}a and the two agree very well to within statistical uncertainties.  This agreement is a further indication of isotropy between $v$ and $w$.

From the autocorrelation curve it is possible to define the Taylor length scale as $\lambda_f=[-\frac{1}{2}f^{\prime\prime}(0)]^{1/2}$. Equivalently, $\lambda_f$  is the point at which the parabola $p$, tangent to $f(r)$ near $r=0$, intersects the axis $r$, so that $p(r)=1+r^2/\lambda_f^2$.  We estimate the coefficients of $p$ by fitting a parabola to the second and third point of $f(r)$. 
The parabolic fit seen in figure \ref{fig:SF2} gives $\lambda_f=15.9$ mm, with a 95\% CI of [14.4 16.4] mm. The uncertainty in this value  is dominated by the small number of data points used in the fit, which is a consequence of the spatial resolution of our PIV measurements.
	
We can use $\lambda_f$ to predict the TKE dissipation rate $\epsilon$, because $\lambda_f$ is related to fluid velocity gradients as follows \citep[see][]{BPope:2000p18029}:
\begin{equation}\label{eq:tay_der}
\left\langle \left(\frac{\partial u}{\partial x}\right)^2 \right\rangle =\frac{2u'^2}{\lambda_f^2}= \frac{4k^2}{3\lambda_f^2}, 
\end{equation}
where $u' \equiv \sqrt{\frac{2}{3}k^2}$. 
For flows in which the strain rate tensor is isotropic, equation \ref{eq:tay_der} can be combined with the definition of $\epsilon$ to give:   
\begin{equation}\label{eq:iso_eps}
\epsilon=15\nu \left\langle \left(\frac{\partial u}{\partial x}\right)^2 \right\rangle=20 \nu \frac{k^2}{\lambda_f^2}.
\end{equation}
Here, $\nu$ is the kinematic viscosity (in our case, for water at 22.8$^\circ$ C, the value is $0.948 \times 10^{-6} $ m$^2$s$^{-1}$). For $\lambda_f=15.9$ mm measured from the autocorrelation function and the value of $k^2$ from table \ref{table:res}, we obtain $\epsilon=4.63 \times 10^{-5}$ m$^2$s$^{-3}$, with a 95\% CI of [4.12 5.14].

The dissipation rate can also be computed in a second way, which will allow us to draw further conclusions about isotropy.  This calculation uses the following result from Kolmogorov theory \citep{Kolmogorov:1941p23204}:
\begin{equation}\label{eq:sf2}
S_L^{(2)}(r) = C_2 \epsilon^{2/3} r^{2/3},
\end{equation}
for $r$-values within the inertial subrange. Although this expression is derived for the case of locally homogeneous and isotropic flows,  experimental evidence \citep{Saddoughi:1994p24184,Sreenivasan:1995p23249} strongly suggests that this holds in heterogeneous flows (\emph{e.g.} boundary layer) and that $C_2=2$. In figure \ref{fig:SF2}b we show the compensated structure function $S_L^{(2)}$.  The value of the plateau at $r > \lambda_f$ gives us a dissipation rate of $ \epsilon = 4.68\times 10^{-5}$ m$^2$s$^{-3}$ with a 95\% CI of [4.45 4.78].  This value of $\epsilon$ is in very good agreement with the value obtained above using the Taylor lengthscale. We use this dissipation rate to determine the Kolmogorov lengthscale as $\eta=(\nu^3/\epsilon)^{1/4}$=0.37 mm, and the Kolmogorov timescale as $\tau_\eta=(\nu/\epsilon)^{1/2}=0.142$ s. 

We use the two--point statistics to evaluate the isotropy across a range of intermediate scales.  
The distance between the measured and predicted $g(r)$ in figure \ref{fig:SF2}a can be seen as a scale--by--scale measure of isotropy.  Furthermore, we can compute $S_L^{(2)}$ along two orthogonal directions ($y$ and $z$) shown in figure~\ref{fig:SF2}b; the distance between these two curves can also be seen as a scale--by--scale measure of isotropy.  Both of these measurement methods suggest that the flow is isotropic at all scales. We note that anisotropy at intermediate and large scales, when present, can be a large source of error in estimating the dissipation rate as done above.  Thus the intermediate-scale isotropy seen in figure 5 and the large-scale isotropy seen in figure 4 support our approaches for estimating the dissipation rate, and the uncertainty in $\epsilon$ is due to statistical fluctuations and not anisotropy.

\begin{figure}
\centering
\includegraphics[width=0.75\textwidth]{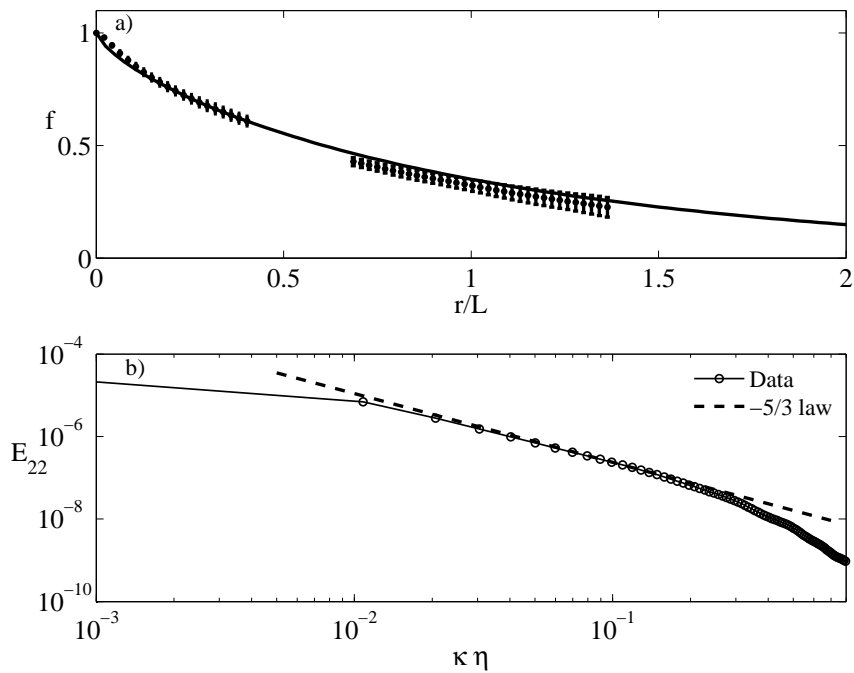}
\caption{a) Longitudinal autocorrelation function, extending over a larger region than in figure \ref{fig:SF2}a.  Symbols indicate the experimental data with 95\% CI, and the solid line shows data interpolation using Bessel function fit performed on the data for $\lambda_f<r<L/2$. b) Longitudinal power spectrum $E_{22}$ computed from the measured autocovariance. The curves are obtained from 1517 independent 2D-PIV measurements.}\label{fig:bessel}  
\end{figure}

Integral lengthscales describe the large scales of turbulence, and thus are useful as a scale with which to evaluate the size of the homogeneous region created  by our stirred tank design.  This is because the homogeneous region at the tank center will only be `truly' homogeneous if its dimensions are significantly larger than all turbulent scales.
We define $L$ as the longitudinal integral length scale, obtained from the integral of the autocorrelation function $f(r)$ from $r=0$ to  $r=\infty$.  Due to practical limitations, it is rare for laboratory measurements to cover a large enough region in space to directly calculate this integral.  That is,  velocity fields must be measured over a length of 3 to 5$L$ before $f(r)$ reaches its asymptotic limit at zero.  In some flows, this limitation can be overcome by assuming space--time equivalence, such as Taylor's frozen turbulence hypothesis.  However, for the flow considered here, spatial and temporal dynamics cannot be translated in a trivial manner.  Thus, to measure $L$ accurately, we combine two strategies.  First, we use a pair of simultaneous 2D-PIV measurements to extend the measurement region in space without losing fine--scale resolution.  Second, we fit a model function to the curve $f(r)$ (see figure \ref{fig:bessel}).  This model is valid for the inertial subrange (\emph{i.e.} where most PIV data is collected) and includes $L$ as one of the two fitting parameters.  This method is preferable to the strategy of extrapolating $f(r)$ until it reaches zero, because no universal model exists for $f(r>L)$ \citep{Davidson_turb}.  Our model of $f(r)$ is obtained from the Kolmogorov hypotheses, specifically by inverse Fourier transform of a power-law model spectrum \citep[see appendix G of][]{BPope:2000p18029}.  This predicts that the autocorrelation curve within the inertial subrange takes the form:
\begin{equation}\label{eq:bess}
f\bigg(\frac{r}{L}\bigg)=\frac{2}{\Gamma(q)}\bigg(\frac{r}{2L}\alpha\bigg)^q K_q\bigg(\frac{r}{L}\alpha\bigg),
\end{equation}
where $\Gamma$ is the gamma function, $K_q$ is the modified Bessel function of the second kind and $\alpha$ is determined by the constraint $f(0)=1$. Equation \ref{eq:bess} has two free parameters: $L$ and $q$, which are the integral length scale and an algebraic restatement of the velocity power spectrum's power law exponent, respectively.  Fitting this model to our data at the tank center gives $L=9.50$ cm (CI $\in$ [9.45 9.55] cm)  and $q=0.353$, which corresponds to a spectral power law exponent of $1.70$ (CI $\in$ [1.700 1.708]).
Although the model is derived to fit data in the inertial subrange, figure \ref{fig:bessel}a shows that the fit also provides a good approximation of the autocorrelation curve for large separations ($r>L/2$). 
The scaling exponent of the power spectrum can be also obtained directly by computing the power spectrum from the inverse Fourier transform of the measured autocovariance. 
Figure \ref{fig:bessel}b shows that the spectrum closely matches a $\kappa^{-5/3}$ slope over about one decade in $\kappa$, which indicates the presence of a well-developed inertial range.   

As a final note on this section, we discuss the relationship between the value of $L$ determined from $f(r)$, and the common approximation $l\equiv \frac{u'^3}{\epsilon}$.  It is currently an open question whether the ratio $\Phi \equiv L/l$ is a constant, a function of the Reynolds number, or even a function of the initial and boundary conditions \citep{Batchelor:1953p24315,Sreenivasan:1984p24353,Gamard:1999p24188}.  We are able to add the result of  $\Phi=0.55$ based on the data presented above.  This agrees well with the model
of \cite{Gamard:1999p24188}, who use scaling arguments to derive weak dependence of $\Phi$ on the Reynolds number (which they define as $Re_G\equiv L/\eta$). They predict $\Phi=0.64$ at $Re_G=275$, and we measure $\Phi=0.55$ at $Re_G=256$, with 95\% CI of [0.50 0.60].  Interestingly, this is very close to the one Reynolds number at which the experimental observations by \cite{Mydlarsky:1996p24448} in grid-generated turbulence disagreed with the theory of Gamard and George.  Thus the support of our measurements is particularly important in evaluating the model.

\section{Spatial variation of turbulent quantities}\label{sec:space}
In the previous section we demonstrated homogeneity and isotropy near the center of our stirred tank. In this section we investigate turbulent statistics over a larger spatial region.  With this, we will assess the full size of the homogeneous and isotropic region in the tank center.  Measurements are done by moving 2D-PIV across four stations in the $z$-direction, thereby measuring over a distance greater than two integral length scales. 

\begin{figure}
\centering
\includegraphics[width=0.85\textwidth]{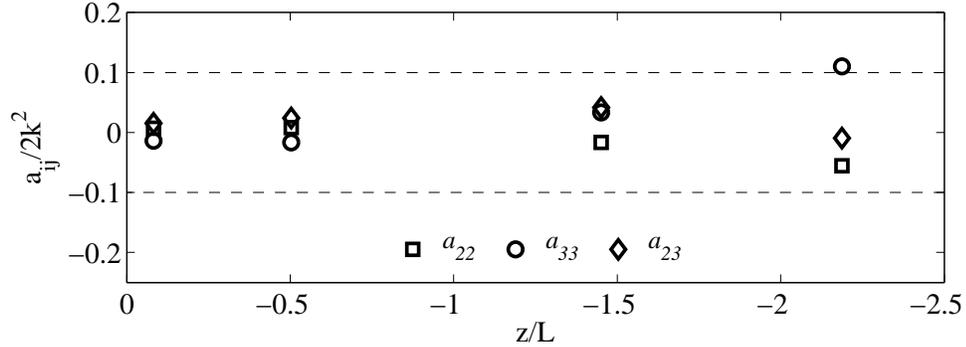}
\caption{Spatial variation of two diagonal ($a_{22}$ and $a_{33}$) and one off-diagonal ($a_{23}$) components of the anisotropy tensor normalized by the local TKE. The dashed lines show the 10\% variation.}\label{fig:aij}  
\end{figure}

\begin{figure}
\centering
\includegraphics[width=0.55\textwidth]{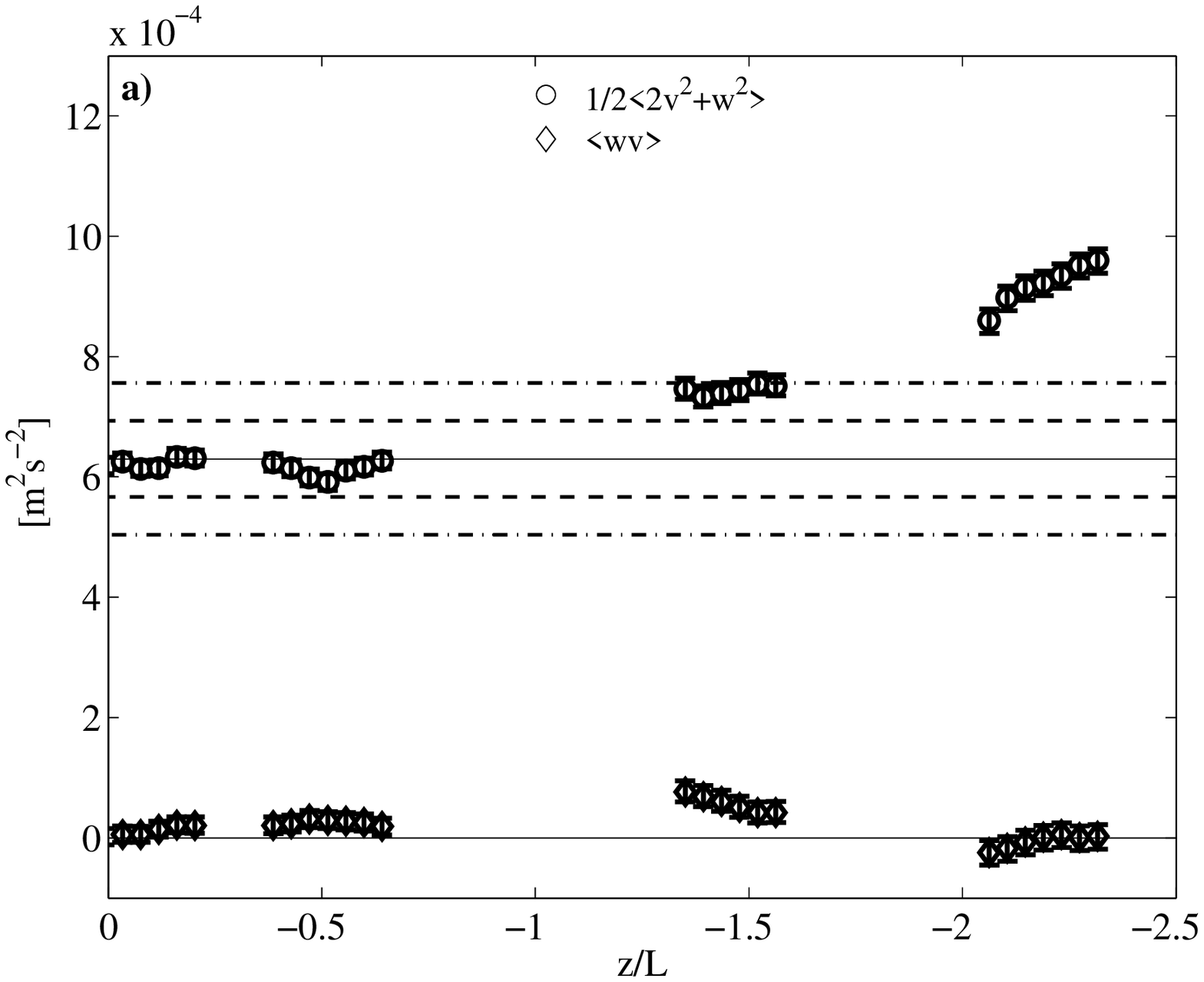}
\includegraphics[width=0.24\textwidth]{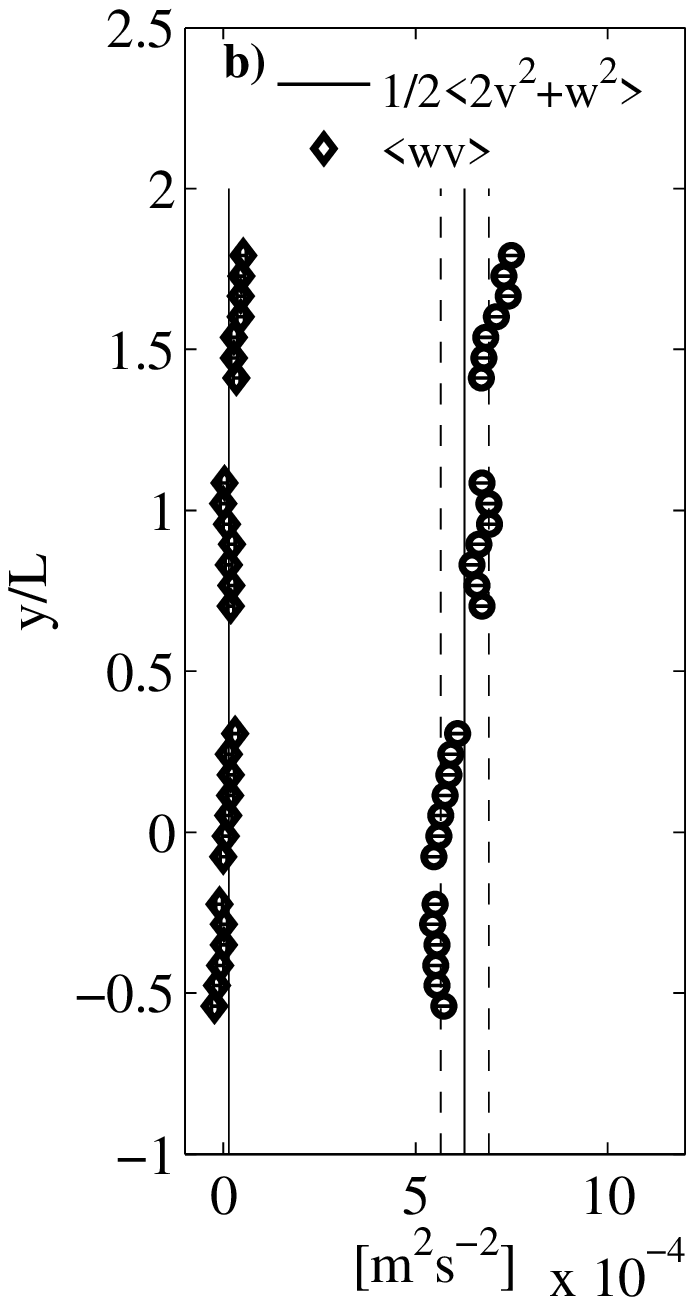}
\caption{Distribution of TKE ($\circ$) and Reynolds stress $\langle vw \rangle$ ($\Diamond$) in the $y-z$ plane.  (a) Data averaged over $y$. (b) Data averaged over $z$. The solid lines indicate mean quantities in the tank center, and percent changes relative to this mean value are shown as dashed (10\% variation) and dash-dot (20\% variation) lines. The errorbars represent the 95\% confidence intervals computed by bootstrap method.}\label{fig:TKEZ}  
\end{figure}

We first analyze one--point statistics, namely  TKE and the Reynolds stress tensor.  
We decompose the Reynolds stress tensor into an isotropic part $\frac{2}{3}k^2\delta_{ij}$, and an anisotropic part $a_{ij}\equiv\langle u_iu_j\rangle-\frac{2}{3}k \delta_{ij}$. The anisotropic part is responsible for the momentum transfer, and it should be zero in a homogeneous flow. 
Given our 2D data, we can compute three components of the anisotropic stress tensor: $a_{22}$, $a_{33}$, and $a_{23}=\langle vw \rangle$. All three components of the anisotropy tensor are seen in Figure \ref{fig:aij}, normalized by the local value of TKE.  This figure shows that in the region $z/L < 1.5$ the anisotropic part of the stress tensor remains well below $10\%$ of the TKE.     
The distribution of TKE and $\langle vw \rangle$ over $y$ and $z$ are shown in figures \ref{fig:TKEZ}a-b. Both figures indicate that TKE varies by less $\ll10\%$ for $0<z/L<1$ and $0.5<y/L< 2$. Furthermore, we see that  $\langle vw \rangle$ is statistically identical to zero, and much smaller than TKE.   Together, the one--point statistics shown in figures \ref{fig:aij} and \ref{fig:TKEZ} strongly suggest that the homogeneous isotropic region extends farther than one integral lengthscale from the tank center (eventual conclusions appear in table \ref{table:sum}).  Before evaluating this further, we examine the two--point statistics.

\begin{figure}
\centering
\includegraphics[width=0.85\textwidth]{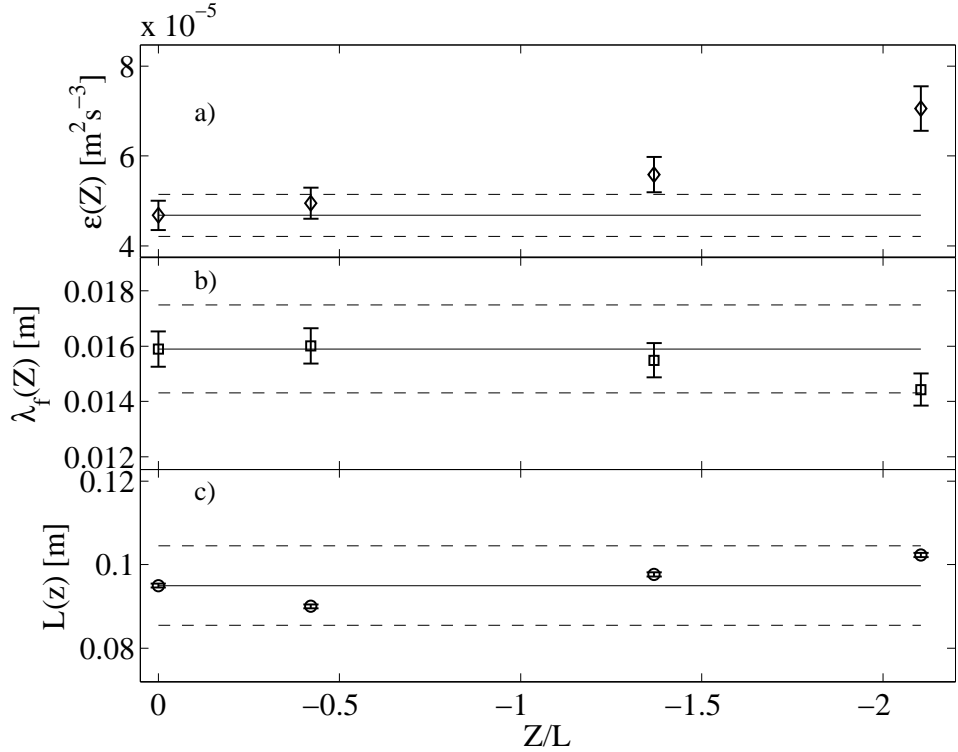}
\caption{Spatial distribution of turbulent quantities within $0\leq z/L \leq 2.5$. The dashed lines show the $\pm 10\%$ variation with respect to the reference value at the tank center (solid line). The errorbars show the 95\% CI computed via bootstrap method.}\label{fig:space}  
\end{figure}

Values of $\epsilon$, $\lambda_f$ and $L$ are computed  at four different locations in $z$, using the methods described in section 5.  The results are shown in figure \ref{fig:space} and table \ref{table:turbz}. Figure \ref{fig:space}a shows that the pattern of the dissipation measurements follows the pattern of  the TKE observed in figure \ref{fig:TKEZ}. Figures \ref{fig:space}b-c show that the turbulent lengthscales stay approximately constant over a wider region than any of the other one-- and two--point statistics reported here.

Given the above results, we conclude that the homogeneous and isotropic region extends at least to $z=-1.0L$.  A conservative quantitative demarcation of the homogeneous region can be made at $z=-1.0L$ using the common $<10\%$ variation criterion.  If we use instead a $< 20\%$ variation criterion the homogeneous region reaches to $z=-1.5L$.  At $z=-1.5L$, most statistical values are within the 10\% demarcation line, and only the dissipation rate and the TKE are clearly trending away from the reference values in a statistically significant manner.  %A common metric used in the literature to demarcate homogeneous regions is the range over which TKE varies by less than 10\%; by this metric, the homogeneous region studied here ends at roughly $z=-1.5L$.

The size of the homogeneous region in $y$ is larger than that in $z$; Figure \ref{fig:TKEZ}b shows that TKE is constant in $y$ at least to $y=2L$.  This behavior is expected from the boundary conditions imposed by the tank's symmetric forcing geometry.  That is, we expect the homogeneous region in $x$ and $y$ to be limited only by the size of the tank.  Results on shear--free turbulence near an interface \citep{hunt_free-stream_1978,perot_shear-free_1995} suggest that the tank wall at $y\approx 4.3L$ will begin to influence the flow at $y\approx2.3L$, and strongly influence it for $y>3.3L$.  
 
Assuming reflective symmetry about the origin and rotational symmetry between lateral velocity components $u$ and $v$, we can use the above results to determine the full size of the homogeneous isotropic region in our stirred tank, which is given in table \ref{table:sum}.
\begin{table}[h] 
\centering
\begin{tabular}{ccccccccc}                      
\hline
\hline
 & $\pm 20\%$  & $\pm 10\%$  \\
 $x$ &       $- 3.3L< x < 3.3L$  &    $- 2.3L< x < 2.3L$  \\
 $y$ &       $- 3.3L< y < 3.3L$   &      $- 2.3L< y < 2.3L$           \\
 $z$ &       $- 1.5L< z < 1.5L$   &    $- 1.0 L< z < 1.0L$             \\
\hline
\hline
 \end{tabular}
 \caption{Two quantitative demarcations of the size of the region at the tank center that contains homogeneous isotropic turbulence.}\label{table:sum}
\end{table}

\begin{table}[tbp] 
\centering
\begin{tabular}{lcccccccc}                      
\hline
\hline
       &             & $\frac{Z}{L}=0$ & $\frac{Z}{L}=-0.42$  & $\frac{Z}{L}=-1.37$ & $\frac{Z}{L}=- 2.1$\\  
 $k^2$ & [$\times10^{-4}$ m$^{2}$ s$^{-2}$] & 6.07  & 5.92  & 6.95 & 8.91\\
&  95\% CI    & [5.50 6.64] & [5.46 6.54] & [6.34 7.57] & [8.07 9.83]\\
 $u'=(\frac{2}{3}k^2)^{\frac{1}{2}}$ &[ $\times10^{-2}$ ms$^{-1}$] & 2.02  & 1.98 & 2.15 & 2.43 \\
 &  95\% CI    & [1.91 2.10] & [1.90 2.08] & [2.05 2.24] & [2.31 2.56]\\
 $v_{rms}/w_{rms}$& - & 0.95 & 0.99  & 0.93 & 0.77 \\
 &  95\% CI    & [0.84 1.09] & [0.89 1.12] & [0.79 1.03] & [0.65 0.89]\\
 \\
 $\epsilon$ & [$\times10^{-5}$ m$^2$s$^{-3}$] & 4.65  & 4.94 &5.58 & 7.05\\
 &  95\% CI    & [4.45 4.78] & [4.68 5.18] & [4.94 6.22] & [6.34 7.76]\\
 $\eta$ & [$\times10^{-3}$ m] & 0.37  & 0.36  & 0.35 & 0.33\\
 &  95\% CI    & [0.36 0.37] & [0.36 0.37] & [0.34 0.36] & [0.32 0.34]\\
 $\lambda_f$ & [$\times10^{-3}$ m] &15.9  & 16.0 & 15.5 & 14.4\\
 &  95\% CI    & [15.4 16.4] & [15.5 16.5] & [15.2 15.9] & [14.0 14.8]\\
 $L$ &[$\times10^{-2}$ m] & 9.50  & 9.00 &  9.70 & 10.20\\
 &  95\% CI    & [9.45 9.55] & [8.95 9.05] & [9.62 9.78] & [10.10 10.30]\\
 \\
 $R_\lambda=\frac{\lambda_fu'}{\nu}$& - & 338 & 334  & 351 & 372 \\
  &  95\% CI    & [310  363] & [304 348] & [328 378] & [344 403]\\
 $Re_L=\frac{u'L}{\nu}$&- & 2002 & 1881  & 2258 & 2617\\
  &  95\% CI    & [1920 2133] & [1819 1991] & [2104 2323] & [2506 2777]\\
\hline
\hline
 \end{tabular}
 \caption{Summary of turbulence quantities}\label{table:turbz}
\end{table}

\section{Conclusions}

Homogeneous isotropic turbulence is of extreme interest for theoretical and engineering problems on turbulent dynamics. However, recreating this idealized situation in a laboratory flow is a formidable challenge due to the local distribution of turbulent production.  We have developed a laboratory flow with an unprecedented degree of homogeneity and isotropy, and with negligible mean flow. The flow is obtained by combining the concepts of randomly actuated synthetic jet arrays \citep{Variano:2008p12422} and symmetric forcing in a stirred tank.   
 
PIV measurements in this flow show that there is a region at the core of the tank in which the flow is homogeneous  over two integral length scales. Several tests of isotropy confirm that the flow in the homogeneous region is isotropic at all scales.  The Reynolds number (based on the Taylor microscale) is between 334 $<$ R$_\lambda$ $<$ 351 in the homogeneous region.  This high value means that we can expect a large inertial subrange, thereby affording a meaningful approximation to turbulence theory.  The mean flow is less than 10\% of the turbulent fluctuating velocity magnitude.  This low mean flow makes it convenient to investigate the dynamics of turbulence because one can independently measure the Eulerian temporal, Eulerian spatial, and Lagrangian statistics at a single, non--moving, measurement location.

A conservative demarcation ($<10\%$ variation of turbulent quantities) of the homogeneous isotropic region is $2.3L>x>-2.3L$, $2.3L>y>-2.3L$, and $1.0L>z>-1.0L$, where $L=9.5$ cm.  A more liberal quantitative assessment  ($<20\%$ variation of turbulent quantities) of the size is $3.3L>x>-3.3L$, $3.3L>y>-3.3L$, and $1.5L>z>-1.5L$.  In either case, this region is much larger than in other laboratory apparati designed to create homogeneous isotropic turbulence at the same high Reynolds number.  The large size of the homogeneous isotropic region is especially important for measuring the dynamics of turbulent particle suspensions.  This is because particle motion depends in part on the history of the flow experienced \cite{Mei:1992p25755}.  Thus, for particle statistics to accurately represent the effects of homogenous isotropic turbulence on particles, they must be measured in a homogeneous region so that they do not include the signature of other regions of the flow.  At the center of the stirred tank discussed here, any particle that is measured will have traveled through a large region of homogeneous turbulence, and thus its motion will be almost entirely due to this flow.

The method of turbulence generation presented here can be extended in a number of possible ways.  Tank geometry can be systematically varied to obtain different turbulent parameters at the tank center, covering a range of Reynolds numbers and dissipation rates.  The large number of jets offers a significant amount of freedom in driving flow patterns, and thus different driving algorithms can be used to tune the mean flow and energy-containing scales.  If it is important for a study, the fraction of the total tank volume that is occupied by HIT can be directly adjusted by extending the lateral tank boundaries.  The method discussed here can also be incorporated in DNS as an alternative forcing mechanism that would likely be better suited for analyzing the effect of turbulence on suspended particles, bubbles, or droplets.

\clearpage
\section*{Acknowledgments}
The authors gratefully acknowledge those who contributed to the design and construction of this facility: CEE staff Joel Carr, Jeff Higginbotham, and Matt Cataleta, and CEE students Colin Meyer, Matt Ritter, Margaret Byron, Jeff Semigran and Laura Mazzaro.

\bibliographystyle{spbasic}
\bibliography{tankarticle2012}

\begin{thebibliography}{48}
\providecommand{\natexlab}[1]{#1}
\providecommand{\url}[1]{{#1}}
\providecommand{\urlprefix}{URL }
\expandafter\ifx\csname urlstyle\endcsname\relax
  \providecommand{\doi}[1]{DOI~\discretionary{}{}{}#1}\else
  \providecommand{\doi}{DOI~\discretionary{}{}{}\begingroup
  \urlstyle{rm}\Url}\fi
\providecommand{\eprint}[2][]{\url{#2}}

\bibitem[{Abdelsamie and Lee(2012)}]{Abdelsamie:2012p24119}
Abdelsamie AH, Lee C (2012) Decaying versus stationary turbulence in
  particle-laden isotropic turbulence: Turbulence modulation mechanism. Phys of
  Fluids 24:015,106

\bibitem[{{\'A}lamo and Jim{\'e}nez(2009)}]{DelAlamo:2009p12425}
{\'A}lamo JCD, Jim{\'e}nez J (2009) Estimation of turbulent convection
  velocities and corrections to taylor's approximation. J Fluid Mech 640:5--26

\bibitem[{Balachandar and Eaton(2010)}]{Balachandar:2010p13976}
Balachandar S, Eaton JK (2010) Turbulent dispersed multiphase flow. Annu Rev of
  Fluid Mech 42:111--133

\bibitem[{Batchelor(1953)}]{Batchelor:1953p24315}
Batchelor G (1953) The theory of homogeneous turbulence. Cambrige Univeristy
  press

\bibitem[{Bellani et~al(2012)Bellani, Byron, Collignon, Meyer, and
  Variano}]{BellaniJFM2012}
Bellani G, Byron ML, Collignon AG, Meyer CR, Variano EA (2012) {Shape effects
  on turbulent modulation by large nearly neutrally buoyant particles}. J Fluid
  Mech 712:41--60

\bibitem[{Benzi et~al(2010)Benzi, Biferale, Fisher, Lamb, and
  Toschi}]{Benzi:2010p23252}
Benzi R, Biferale L, Fisher R, Lamb DQ, Toschi F (2010) Inertial range eulerian
  and lagrangian statistics from numerical simulations of isotropic turbulence.
  J Fluid Mech 653:221--244

\bibitem[{Birouk et~al(2003)Birouk, Sarh, and G{\"o}kalp}]{Birouk:2003p23787}
Birouk M, Sarh B, G{\"o}kalp I (2003) An attempt to realize experimental
  isotropic turbulence at low reynolds number. Flow, Turbulence and Combustion
  70:325--348

\bibitem[{Chang et~al(2012)Chang, Bewley, and Bodenschatz}]{Chang2012}
Chang K, Bewley GP, Bodenschatz E (2012) {Experimental study of the influence
  of anisotropy on the inertial scales of turbulence}. J Fluid Mech
  692:464--481

\bibitem[{Davidson(2004)}]{Davidson_turb}
Davidson PA (2004) {Turbulence: An Introduction for Scientists and Engineers}.
  Oxford

\bibitem[{Dennis and Nickels(2008)}]{Dennis:2008p12424}
Dennis DJC, Nickels TB (2008) On the limitations of taylor's hypothesis in
  constructing long structures in a turbulent boundary layer. J Fluid Mech
  614:197--206

\bibitem[{Douady et~al(1991)Douady, Couder, and Brachet}]{Douady:1991p23522}
Douady S, Couder Y, Brachet ME (1991) Direct observation of the intermittency
  of intense vorticity filaments in turbulence. PRL 67(8):983--986

\bibitem[{Efron and Tibshirani(1994)}]{efron_introduction_1994}
Efron B, Tibshirani R (1994) An Introduction to the Bootstrap, 1st edn. Chapman
  \& {Hall/CRC}

\bibitem[{Galanti and Tsinober(2000)}]{Galanti:2000p23995}
Galanti B, Tsinober A (2000) Self-amplification of the field of velocity
  derivatives in quasi-isotropic turbulence. Phys Fluids 12(12):3097--3099

\bibitem[{Gamard and George(1999)}]{Gamard:1999p24188}
Gamard S, George WK (1999) Reynolds number dependence of energy spectra in the
  overlap region of isotropic turbulence. Flow, Turbulence and Combustion
  63:443--4777

\bibitem[{George and Davidson(2004)}]{George:2004p24571}
George W, Davidson L (2004) Role of initial conditions in establishing
  asymptotic flow behavior. AIAA J 42(3):438--446

\bibitem[{George(1992)}]{George:1992p24570}
George WK (1992) The decay of homogeneous isotropic turbulence. Phys Fluids
  4(7):1492--1509

\bibitem[{Goepfert et~al(2010)Goepfert, Mari{\'e}, Chareyron, and
  Lance}]{Goepfert:2010p26074}
Goepfert C, Mari{\'e} J, Chareyron D, Lance M (2010) Characterization of a
  system generating a homogeneous isotropic turbulence field by free synthetic
  jets. Exp Fluids 48:809--822

\bibitem[{Guala et~al(2008)Guala, Liberzon, Hoyer, and
  Tsinober}]{Guala:2008p14427}
Guala M, Liberzon A, Hoyer K, Tsinober A (2008) Experimental study on
  clustering of large particles in homogeneous turbulent flow. Journal of
  Turbulence 9(34):1--20

\bibitem[{Hunt and Graham(1978)}]{hunt_free-stream_1978}
Hunt JCR, Graham JMR (1978) {Free-Stream} turbulence near plane boundaries. J
  Fluid Mech 84(02):209--235

\bibitem[{Hwang and Eaton(2004)}]{Hwang:2004p23470}
Hwang W, Eaton JK (2004) Creating homogeneous and isotropic turbulence without
  a mean flow. Exp Fluids 36:444--454

\bibitem[{Kolmogorov(1941)}]{Kolmogorov:1941p23204}
Kolmogorov A (1941) The local structure of turbulence in incompressible viscous
  fluid for very large reynolds numbers. Dokl Akad Nauk SSSR 30:299--303

\bibitem[{Krawczynski et~al(2010)Krawczynski, Renou, and
  Danaila}]{Krawczynski:2010p26179}
Krawczynski JF, Renou B, Danaila L (2010) {The structure of the velocity field
  in a confined flow driven by an array of opposed jets}. Phys Fluids
  22:045,104

\bibitem[{Krogstad and Davidson(2012)}]{Krogstad:2012p23266}
Krogstad PA, Davidson PA (2012) Near-field investigation of turbulence produced
  by multi-scale grids. Phys Fluids 24:035,103

\bibitem[{Kurian and Fransson(2009)}]{Kurian:2009p24110}
Kurian T, Fransson J (2009) Grid-generated turbulence revisited. Fluid Dynamics
  Research 41:021,403(32pp)

\bibitem[{Liu et~al(1999)Liu, Katz, and Menevau}]{Liu:1999p23761}
Liu S, Katz J, Menevau C (1999) Evolution and modelling of subgrid scales
  during rapid straining of turbulence. J Fluid Mech 387:281--320

\bibitem[{Lucci et~al(2010)Lucci, Ferrante, and Elghobashi}]{Lucci:2010p15399}
Lucci F, Ferrante A, Elghobashi S (2010) Modulation of isotropic turbulence by
  particles of taylor length-scale size. J Fluid Mech 650:5--55

\bibitem[{Mei(1992)}]{Mei:1992p25755}
Mei R (1992) History force on a sphere due to a step change in the free-stream
  velocity. Int J Multiphase flow 19(3):509--525

\bibitem[{Moin(2009)}]{Moin:2009p12426}
Moin P (2009) Revisiting taylor's hypothesis. J Fluid Mech , Focus on Fluids
  640:1--4

\bibitem[{Mydlarski and Warhaft(1996)}]{Mydlarsky:1996p24448}
Mydlarski L, Warhaft Z (1996) On the onset of high-reynolds-number
  grid-generated wind tunnel turbulence. J Fluid Mech 320:331--368

\bibitem[{Perot and Moin(1995)}]{perot_shear-free_1995}
Perot B, Moin P (1995) {Shear-Free} turbulent boundary layers. part 1. physical
  insights into {Near-Wall} turbulence. J Fluid Mech 295:199--227

\bibitem[{Poelma and Ooms(2006)}]{Poelma:2006p15458}
Poelma C, Ooms G (2006) Particle-turbulence interaction in a homogeneous,
  isotropic turbulent suspension. Applied Mechanics Reviews 59:78--89

\bibitem[{Pope(2000)}]{BPope:2000p18029}
Pope SB (2000) Turbulent flows. Cambridge University Press, 2000

\bibitem[{Raffel et~al(2001)Raffel, Willert, Wereley, and
  Kompenhans}]{RaffelPIV}
Raffel M, Willert CE, Wereley ST, Kompenhans J (2001) {Particle image
  velocimetry}. Springer

\bibitem[{Saarenrinne et~al(2001)Saarenrinne, Piirto, and
  Eloranta}]{Saarenrinne:2001gx}
Saarenrinne P, Piirto M, Eloranta H (2001) {Experiences of turbulence
  measurement with PIV}. Meas Sci Technol 12:1904--1910

\bibitem[{Saddoughi and Veeravalli(1994)}]{Saddoughi:1994p24184}
Saddoughi SG, Veeravalli SV (1994) Local isotropy in turbulent boundary layers
  at high reynolds number. J Fluid Mech 268:333--372

\bibitem[{Shy et~al(1997)Shy, Tang, and Fann}]{Shy:1997p23720}
Shy SS, Tang CY, Fann SY (1997) A nearly isotropic turbulence generated by a
  pair of vibrating grids. Exp Thermal and Fluid Sci 14:251--262

\bibitem[{Srdic et~al(1996)Srdic, Fernando, and Montenegro}]{Srdic:1996p23573}
Srdic A, Fernando HJS, Montenegro L (1996) Generation of nearly isotropic
  turbulence using two oscillating grids. Exp Fluids 20:395--397

\bibitem[{Sreenivasan(1984)}]{Sreenivasan:1984p24353}
Sreenivasan KR (1984) On the scaling of the turbulence energy dissipation rate.
  Phys Fluids 27(5):1048--1050

\bibitem[{Sreenivasan(1995)}]{Sreenivasan:1995p23249}
Sreenivasan KR (1995) On the universality of the kolmogorov constant. Phys
  Fluids 7(11):1--7

\bibitem[{Stanislas et~al(2008)Stanislas, Okamoto, K{\"a}hler, and
  Westerweel}]{Stanislas:2008p9497}
Stanislas M, Okamoto K, K{\"a}hler C, Westerweel J (2008) Main results of the
  third international piv challenge. Exp Fluids 45:27--71

\bibitem[{Talamelli et~al(2009)Talamelli, Persiani, Fransson, Alfredsson,
  Johansson, Nagib, R{\"u}edi, Sreenivasan, and Monkewitz}]{superpipe}
Talamelli A, Persiani F, Fransson JHM, Alfredsson PH, Johansson AV, Nagib HM,
  R{\"u}edi JD, Sreenivasan KR, Monkewitz PA (2009) {CICLoPE---a response to
  the need for high Reynolds number experiments}. Fluid Dyn Res 41:021,407

\bibitem[{Toschi and Bodenschatz(2009)}]{Toschi:2009p13919}
Toschi F, Bodenschatz E (2009) Lagrangian properties of particles in
  turbulence. Annual Reviews 41:375--404

\bibitem[{Tsinober(2004)}]{Tsinober:2004p17930}
Tsinober A (2004) An informal introduction to turbulence. Kluwer academic
  publisher

\bibitem[{Variano and Cowen(2008)}]{Variano:2008p12422}
Variano EA, Cowen EA (2008) A random-jet-stirred turbulence tank. J Fluid Mech
  604:1--32

\bibitem[{Variano et~al(2004)Variano, Bodenschatz, and
  Cowen}]{Variano:2004p23918}
Variano EA, Bodenschatz E, Cowen EA (2004) A random synthetic jet array driven
  turbulence tank. Exp Fluids 37:613--615

\bibitem[{Villermaux et~al(1995)Villermaux, Sixou, and
  Gagne}]{Villermaux:1995p23911}
Villermaux E, Sixou B, Gagne Y (1995) Intense vortical structures in
  grid‐generated turbulence. Phys Fluids 7(8):2008--2013

\bibitem[{Voth et~al(2002)Voth, Porta, and Crawford}]{Voth:2002p14090}
Voth G, Porta AL, Crawford A (2002) Measurement of particle accelerations in
  fully developed turbulence. J Fluid Mech 469:121--160

\bibitem[{Zimmermann et~al(2010)Zimmermann, Xu, Gasteuil, Bourgoin, Volk,
  Pinton, and Bodenschatz}]{zimmerman2010}
Zimmermann R, Xu H, Gasteuil Y, Bourgoin M, Volk R, Pinton JF, Bodenschatz E
  (2010) The lagrangian exploration module: an apparatus for the study of
  statistically homogeneous and isotropic turbulence. Rev Sci Instru 81:055,112

\end{thebibliography}

\end{document}